Transport Properties of Ni, Co, Fe, Mn Doped $Cu_{0.01}Bi_2Te_{2.7}Se_{0.3}$ for Thermoelectric Device Applications

*K. C. Lukas, W. S. Liu, Z. F. Ren, C. P. Opeil*
*Department of Physics, Boston College, Chestnut Hill, Massachusetts 02467, USA*

ABSTRACT

$Bi_2Te_3$ based thermoelectric devices typically use a nickel layer as a diffusion barrier to block the diffusion of solder or copper atoms from the electrode into the thermoelectric material. Previous studies have shown degradation in the efficiency of these thermoelectric devices may be due to the diffusion of the barrier layer into the thermoelectric material. In this work Ni, Co, Fe, and Mn are intentionally doped into $Cu_{0.01}Bi_2Te_{2.7}Se_{0.3}$ in order to understand their effects on the thermoelectric material. Thermoelectric transport properties including the Seebeck coefficient, thermal conductivity, electrical resistivity, carrier concentration, and carrier mobility of $Cu_{0.01}Bi_2Te_{2.7}Se_{0.3}$ doped with 2 atomic percent M (M=Ni, Co, Fe, Mn) as $Cu_{0.01}Bi_2Te_{2.7}Se_{0.3}M_{0.02}$, are studied in a temperature range of 5-525 K.

INTRODUCTION

Much work has been done in recent years in an effort to improve the thermoelectric (TE) transport properties of several thermoelectric materials where efforts have specifically focused on the enhancement of the dimensionless figure of merit, $ZT$ where $ZT = (S^2/\rho\kappa)T$ with $S$ being the Seebeck coefficient, $\rho$ the resistivity, $\kappa$ the thermal conductivity, and $T$ the absolute temperature [1]. Efforts have predominantly focused on the reduction of the lattice thermal conductivity through various nano approaches [2, 3]. Ultimately however the material is going to be used in an actual device, where it is necessary for further optimization beyond that of the TE material used. There have been previous studies on the fabrication of TE devices for cooling and power generation, along with the difficulties that arise during fabrication and operation [2]. TE devices are typically constructed with several p-n couples connected electrically in series and thermally in parallel as described by Snyder, *et al*. [4]. A simple schematic of one p-n couple is shown in Figure 1. One of the most important obstacles is to ensure that there is excellent contact between the heating or cooling surface and the p-n couples. Optimizing the contacts benefits the TE device in two ways. First, any electrical contact resistance can be minimized thereby negating effects of Joule heating and thereby increasing the efficiency of the device. Second, the contacts should minimize any thermal contact resistance so that the largest temperature gradient possible can be established; the efficiency of TE materials increases with both *ZT* as well as the temperature gradient [1,2].

The best way to create good electrical and thermal connections is by soldering. However, it is known that typical solders as well as Cu metal (used for electrodes) readily diffuse into and degrade the properties of thermoelectric materials, specifically $Bi_2Te_3$ based materials [2-9]. In order to impede the diffusion of solder into the TE material, diffusion barriers are used [5]. For example, Fe has been attempted as a diffusion barrier in TAGS-85 and PbTe [8,9]. These diffusion barriers are typically thin sputtered or electrochemical deposited metallic elements to again minimize any thermal or electrical contact resistance, and the elements will also ideally have similar coefficients of thermal expansion (CTE) to ensure the mechanical longevity of the device [2,7].

It was previously demonstrated that nickel could be used successfully as a diffusion barrier for solder in p-type $Bi_2(SbTe)_3$ and n-type $Bi_2(TeSe)_3$ alloys [5]. The problem was that even though the nickel stopped the solder, the nickel itself diffused into the n-type $Bi_2(TeSe)_3$. A simple schematic of the Ni diffusion described in ref. 5 in one p-n couple is shown in Figure 1. Shown on the left is the ideal scenario where the Ni acts as a diffusion barrier without diffusing into the TE material. On the right is what actually happens during fabrication and operation. The Ni does not diffuse into the p-type element but does diffuse slightly into the n-type element. A later study demonstrated that using Co as a diffusion barrier prevented the solder from getting into the n-type $Bi_2(TeSe)_3$ while the Co itself did not diffuse into the material as readily as Ni [7]. Each study included only interfacial microstructure data and therefore it is unclear how the diffused Ni or Co affected the transport properties, though it was noted that Peltier cooling devices degraded over time [5]. TE device degradation was attributed to diffusion of the Ni barrier into the material where the device performed worse as time went on. One possible explanation given was that as the Ni diffused further into the TE material [5], the efficiency went down, and so when the Ni fully diffused into the TE material, the device should operate at its lowest efficiency.

In this study we intentionally dope two atomic percent Ni, Co, Fe, and Mn into $Cu_{0.01}Bi_2Te_{2.7}Se_{0.3}$ to study the maximum degradation in electrical and thermal transport properties that a TE material can undergo if these elements are used as a diffusion barrier. Based on a typical device height of 1 mm as well as the thickness of the applied diffusion barrier of 3 μm, the amount of Ni, Co, Fe, or Mn that diffuses into the TE material should not exceed 2 percent [5]. Ni, Co, and Fe all have similar coefficients of thermal expansion which are reported to be similar to that of $Bi_2(TeSe)_3$ [7,10], and therefore would be ideal for device fabrication. Mn has a coefficient of thermal expansion roughly double that of Ni, Co, or Fe but is included in this study to see how transport is effected [10].

## EXPERIMENTAL

Proper stoichiometric amounts of Cu (Alfa Aesar 99.999%), Bi (Alfa Aesar 99.999%), Te (Alfa Aesar 99.999%), Se (Alfa Aesar 99.999%), Ni (Alfa Aesar 99.999%), Co (Alfa Aesar 99.999%), Fe (Alfa Aesar 99.999%), and Mn (Alfa Aesar 99.999%) were prepared by ball milling and hot pressing methods described previously according to the formula $Cu_{0.01}Bi_2Te_{2.7}Se_{0.3}M_{0.02}$ (M=Ni, Co, Fe and Mn) [11]. Thermal conductivity $\kappa$, electrical resistivity $\rho$, Seebeck coefficient $S$, and Hall coefficient $R_H$ were measured using a Physical Properties Measurement System (PPMS) from Quantum Design in a temperature range of 5-350 K. The carrier concentration $n$ and mobility $\mu_H$ were obtained from Hall measurements and the relations $n = 1/R_H q$ and $\mu_H = R_H/\rho$ where $q$ is the electronic charge. Values for the electrical resistivity and Seebeck coefficient at temperatures of 300-525 K were made using both a ZEM-3 from Ulvac Inc. as well as a homebuilt system. For clarity only values from the ZEM-3 are shown, but all data agrees within experimental error. Values for thermal conductivity were obtained using a LaserFlash system from Nietzche. Low temperature measurements of $\kappa$, $\rho$, and $S$ were made on samples parallel to the press direction, while hall measurements as well as high temperature measurements were made perpendicular to the press direction. It has been previously demonstrated that these materials are isotropic to within 10% [11]. Estimated errors for $\rho$, $\kappa$, $S$, $ZT$, $n$, and $\mu_H$ should not exceed 3, 8, 5, 14, 10, and 10%, respectively.

RESULTS AND DISCUSSION

The Hall Coefficient is negative over the entire temperature range showing that the majority carriers are electrons as is to be expected for $Cu_{0.01}Bi_2Te_{2.7}Se_{0.3}$ [11]. Figure 2a shows the carrier concentration for all samples from 10 – 350 K. In contrast to $Cu_{0.01}Bi_2Te_{2.7}Se_{0.3}$ ($4.29\times10^{19}$ /cm$^{-3}$ at 300K), the carrier concentration at room temperature of $Cu_{0.01}Bi_2Te_{2.7}Se_{0.3}M_{0.02}$ is increased with the addition of Ni ($5.72\times10^{19}$ /cm$^{-3}$), Co ($4.88\times10^{19}$ /cm$^{-3}$), and Fe ($4.83\times10^{19}$ /cm$^{-3}$), while $n$ decreased with Mn ($3.11\times10^{19}$ /cm$^{-3}$). The increased carrier concentration for the Ni, Co and Fe, could be explained in the same manner as the introduction of Cu to n-type $Bi_2Te_{2.7}Se_{0.3}$. It was previously shown that Cu easily diffuses into $Bi_2Te_3$ through the interstitial sites between two Te layers and acts as a strong donor providing 0.3 carrier/atom [11]. Ni, Co and Fe appear to get into the interstitial site of $Bi_2Te_{2.7}Se_{0.3}$ as well and also work as an n-type dopant providing 0.12, 0.05, and 0.04 carrier/atom, respectively. Such transition metals have also demonstrated n-type doping behavior when they are located at the interstitial site of layered compounds $TiSe_2$ [12] and $TiS_2$ [13]. The inability of Ni, Co, and Fe to donate an equivalent number of free electrons in $Cu_{0.01}Bi_2Te_{2.7}Se_{0.3}M_{0.02}$, may be associated with the reduced number of outer valence electrons. It seems that Mn did not get into the interstitial site, but substituted for Bi which is known to behave as an acceptor. Figure 2b plots the mobility against temperature up to 350 K. It can be seen that the variation in mobility is quite small with the greatest change being roughly a 10% decrease in $\mu_H$ for the Fe doped sample.

The electrical resistivity, thermal conductivity, Seebeck coefficient and $ZT$ are plotted in Figure 3 from 5-350 K. All thermal and electrical transport properties show a small variation with the introduction of any of the doping impurities. The electrical resistivity, Fig. 3a, shows metallic like behavior. The addition of Ni slightly decreases the resistivity while the addition of Co, Fe, and Mn slightly increases values for $\rho$ where Mn shows the largest increase which is due to a drop in carrier concentration. The Seebeck coefficient is negative over the entire temperature range confirming the majority carriers are electrons, and $S$ is slightly decreased by the addition of Ni. Co and Fe do not appear to impact values for $S$, while the addition of Mn slightly increases $S$ again due to the decrease in $n$. Thermal conductivity decreases with the introduction of either Ni, Co, Fe, or Mn, all of which act as a scattering mechanism as has been previously demonstrated with Cu in $Bi_2Te_{2.7}Se_{0.3}$ [11]. The values of *ZT* seen in Fig. 3d show that overall the figure of merit is unaffected by any of the above 3d transition metal doping.

The electrical resistivity, Seebeck coefficient, thermal conductivity and *ZT* are plotted in Figure 4 from 300-525 K. Again the inclusion of Ni gives a lower value for the electrical resistivity while Mn increases the resistivity. Within experimental error the Seebeck coefficient remains the same except for the Ni doped sample which is suppressed due to the increase in carrier concentration. At these higher temperatures phonon-phonon scattering becomes dominant and all samples exhibit similar values for thermal conductivity, as expected. Overall, in each sample, an increase in $\rho$ comes with an increase in S and therefore all samples have similar values for *ZT* just as in the low temperature data presented in Figure 3. The transport properties measured by the two commercial systems along different pressing directions match to within about 10% as is expected [11].

Both high and low temperature measurements made perpendicular and parallel to the press direction show similar quantitative trends due to the introduction of Ni, Fe, Co, and Mn, where overall there is negligible change to *ZT* with the addition of impurities. This means that if any of these elements are used as diffusion barriers, and diffuse into the TE element there should be negligible effects on the efficiency of the device. It was mentioned previously that former studies showed a decrease in device efficiency which was attributed to diffusion of Ni into the TE material [5]. However, these results are contrary to that understanding because the device degradation is not due to deteriorating TE material performance. One possible explanation is that there is Cu already introduced into the $Bi_2Te_{2.7}Se_{0.3}$ material in this study. Previous studies by Liu, *et. al* added Cu in order to make the material properties reproducible [11]. And therefore with a further introduction of impurities at such a small percentage, there is not as drastic a change in the thermal or electrical transport properties as what is seen with the addition of slight amounts of Cu. Based on this understanding, any of the metallic elements, Ni, Co, Fe, and Mn, can be used as a diffusion barrier for $Cu_{0.01}Bi_2Te_{2.7}Se_{0.3}$ as long as they inhibit the solder from entering the TE material and have the proper mechanical properties during operation [2]. It would be interesting to create a device using $Cu_{0.01}Bi_2Te_{2.7}Se_{0.3}$ as the n-type TE material and Ni as a diffusion barrier. Based on the above data, the device should not degrade if a small amount of Ni diffuses into the TE material as long as there is no diffusion in of the solder. If the device does degrade, then there are other issues with the device. Perhaps some special type of failure at the interface happens during operation, but this failure would not be due to TE material degradation.

The above analysis only applies to bulk or “macro” TE devices, however the data can be useful in the construction of “micro” TE devices as well. Microelectromechanical systems (MEMS) and other thin film TE devices are also of interest and have become a widely studied area [2,4,14-16]. In thin films the diffusion barrier thickness is on the same order as the TE material [2,4], and so if Ni readily diffuses into $Cu_{0.01}Bi_2Te_{2.7}Se_{0.3}$ then the inclusion of Ni will be greater than the 2% addition studied here. However previous studies for Co show that it does not easily diffuse into $Bi_2(TeSe)_3$ and therefore the low doping percentage study here can be applicable [7]. Co does go through a structural phase transition at 380 $^{o}$C [17], but this is higher than the typical operating temperature of $Bi_2(TeSe)_3$ based TE devices and therefore should be of no concern [2,5].

Due to the similarity in the coefficient of thermal expansion, minimal diffusion of Co into $Bi_2(TeSe)_3$ [7], and no significant change in *ZT* for Co doped $Cu_{0.01}Bi_2Te_{2.7}Se_{0.3}$, it appears that Co would be an excellent choice for the contact material in a $Cu_{0.01}Bi_2Te_{2.7}Se_{0.3}$ thin film TE device. The authors could find no information on the diffusion of Fe or Mn in $Bi_2(TeSe)_3$, however if they show similar diffusion tendencies to that of Co then they could also be a possible option for a contact material. However, because Fe oxidizes easily and Mn has a higher CTE, Co appears to be the ideal choice for contact materials in either macro or micro TE devices.

## CONCLUSION

Thermoelectric transport properties of $Cu_{0.01}Bi_2Te_{2.7}Se_{0.3}$ doped with 2 atomic percent Ni, Co, Fe, and Mn are studied to reveal information on possible metallic elements for use as diffusion barriers in TE devices. It is shown that *ZT* is unaffected by the low percentage impurity doping and therefore the efficiency of TE devices should not be affected if any of these metals diffuse into the TE material while being used as a diffusion barrier for solder. And based on this and previous studies Co seems to be the optimal choice for a diffusion barrier. It is also noted that the addition of Cu into $Bi_2Te_{2.7}Se_{0.3}$ could be of great benefit to TE device fabrication because any excess introduction of impurities from the diffusion barriers would be negligible, if it is in fact the impurities that are degrading the performance.

## ACKNOWLEDGEMENTS

We gratefully acknowledge funding for this work through the “Solid State Solar-Thermal energy conversion Center (S3TEC)”, an Energy Frontier research Center founded by the U.S. Department of Energy, Office of Basic Energy Science under award number: DE-SC0001299/DE-FG02-09ER46577.

List of Captions

Figure 1: Both (a) and (b) show one p-n couple that can be used as a thermoelectric generator or Peltier cooler. Typical TE devices contain many couples. The ideal solder connection to the Ni diffusion barrier for a $Bi_2Te_3$ p-n couple shown in (a) where the Ni does not diffuse into the TE material. (b) shows the actual case during fabrication and operation where the Ni diffuses into the n-type TE material which is believed to be the cause of device degradation [4].

Figure 2: The carrier mobility (a) and carrier concentration (b) are plotted with temperature from 5-350 K.

Figure 3: The electrical resistivity (a), thermal conductivity (b), Seebeck coefficient (c), and *ZT* (d) are plotted with temperature from 5-300 K.

Figure 4: The electrical resistivity (a), thermal conductivity (b), Seebeck coefficient (c), and *ZT* (d) are plotted with temperature from 300-525 K.

Figure 1

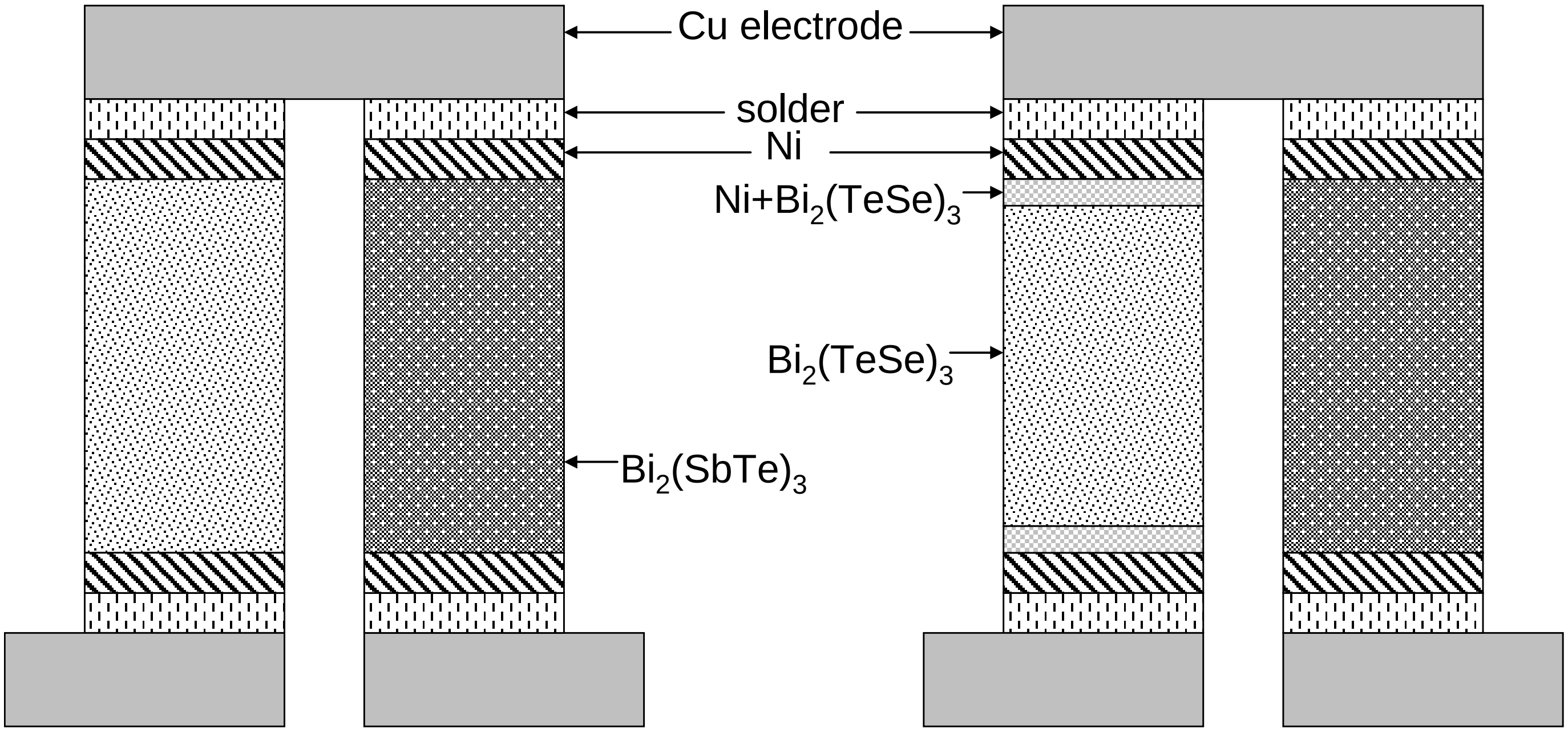

Figure 2

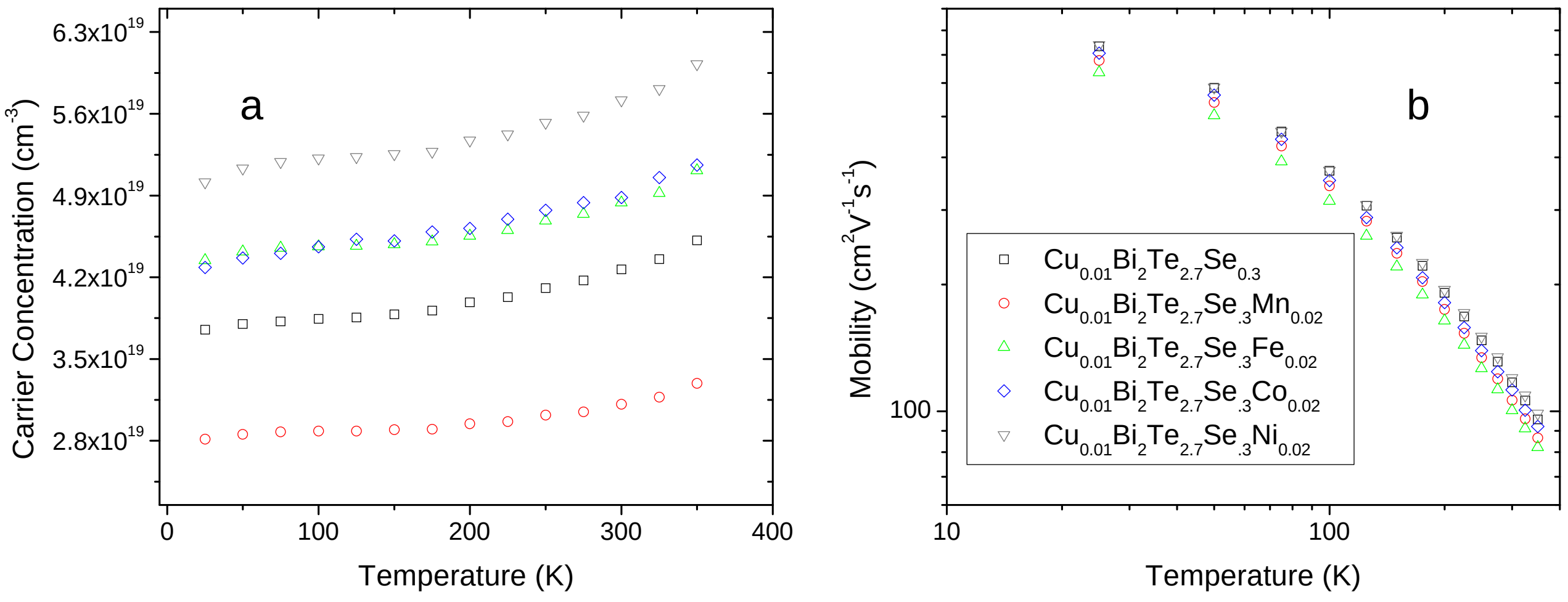

Figure 3

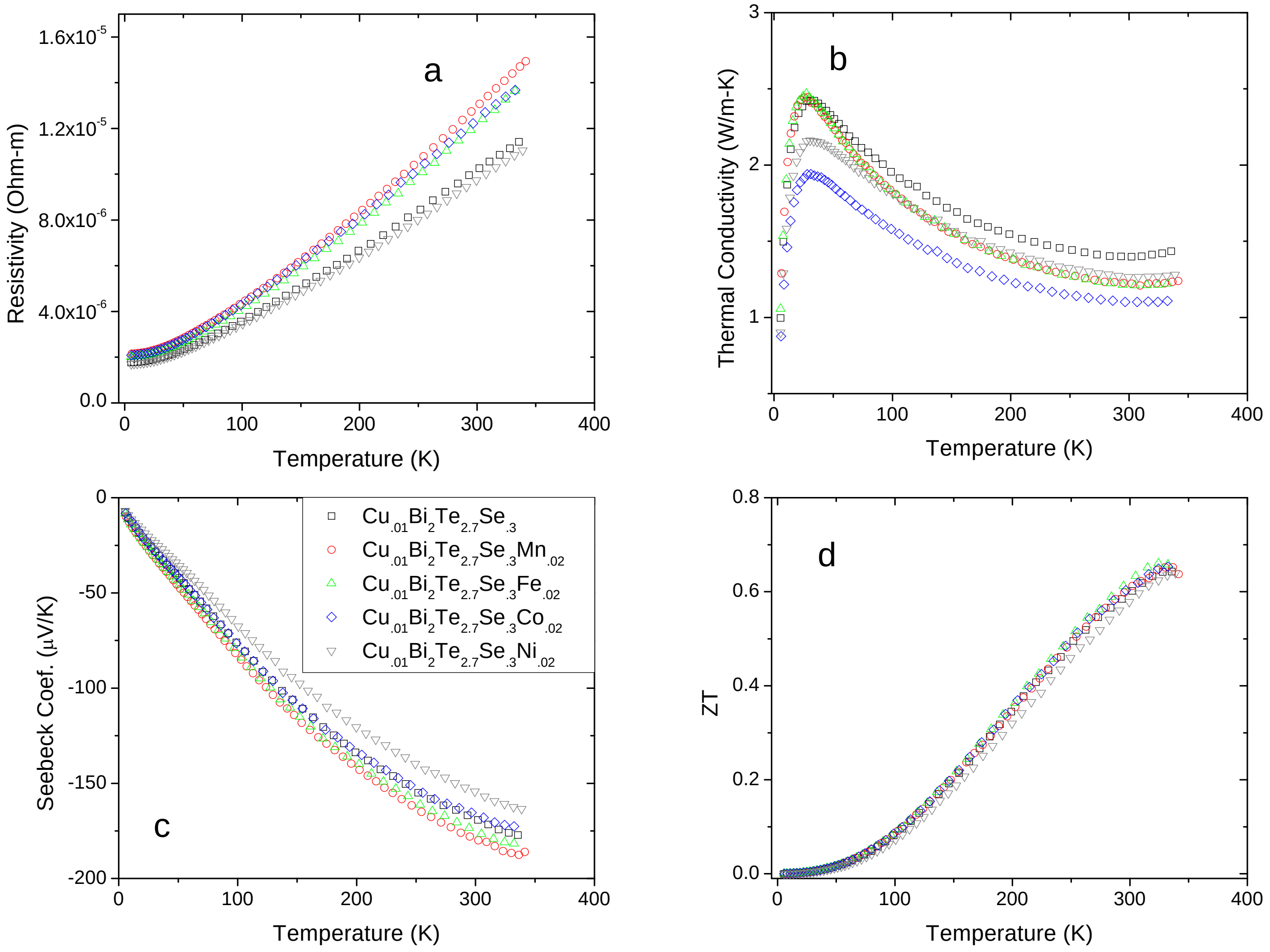

Figure 4

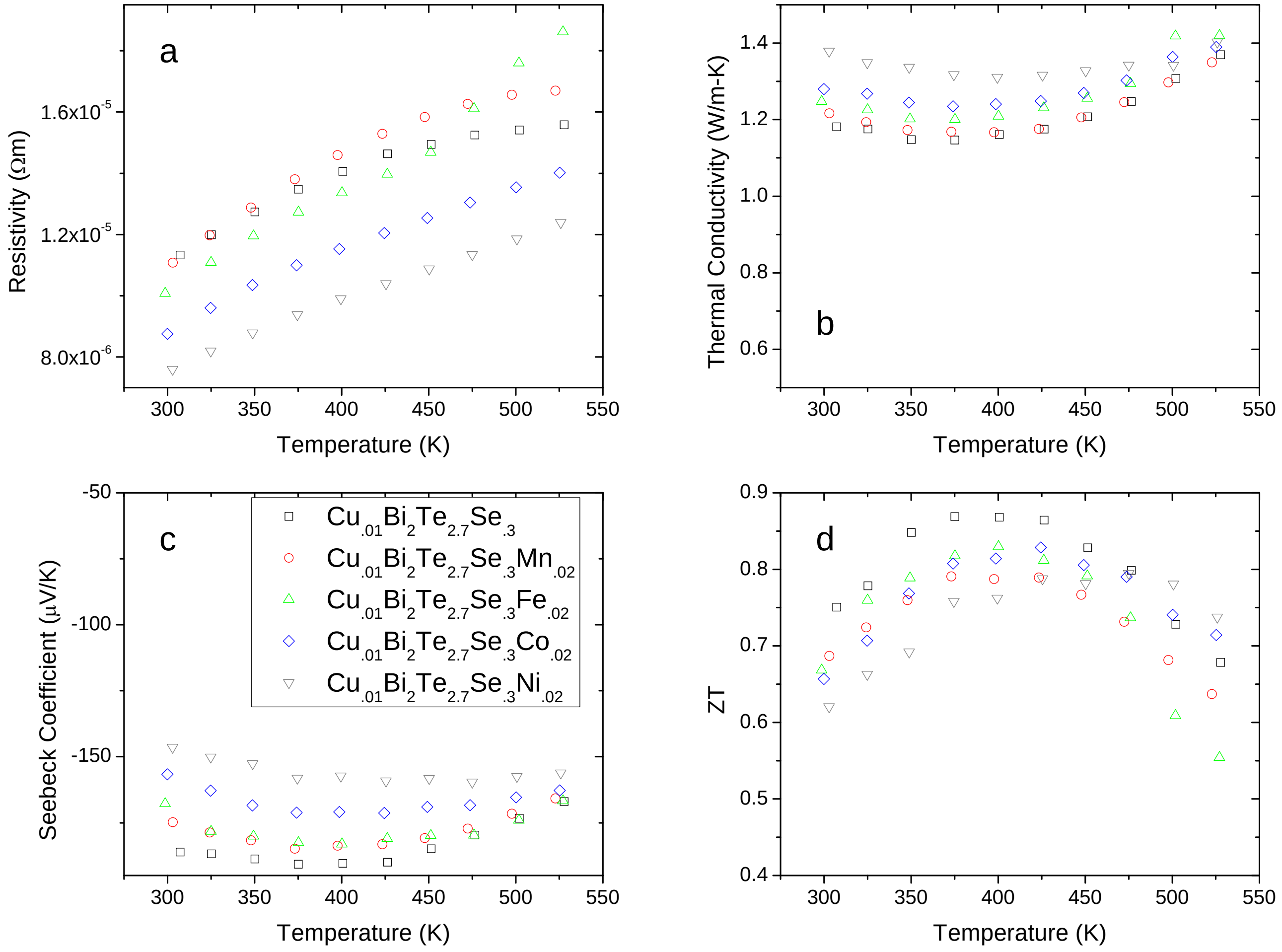